\documentclass[longauth]{aa}
\usepackage{graphicx}
\usepackage{txfonts}
\usepackage{xcolor}
\usepackage{hyperref}
\hypersetup{colorlinks, linkcolor=blue, citecolor=blue, urlcolor=blue}

\newcommand{\pivec}{\mbox{\boldmath $\pi$}}
\newcommand{\muvec}{\mbox{\boldmath $\mu$}}
\newcommand{\xivec}{\mbox{\boldmath $\xi$}}

\newcommand{\te}{t_{\rm E}}
\newcommand{\thetae}{\theta_{\rm E}}

\newcommand{\pie}{\pi_{\rm E}}
\newcommand{\pien}{\pi_{{\rm E},N}}
\newcommand{\piee}{\pi_{{\rm E},E}}
\newcommand{\dl}{D_{\rm L}}
\newcommand{\ds}{D_{\rm S}}
\def\e{{\rm E}}

\definecolor{brown}{rgb}{0.59, 0.29, 0.0}
\definecolor{darkgreen}{rgb}{0.0, 0.42, 0.24}
\definecolor{darkblue}{rgb}{0.01, 0.31, 0.59}
\definecolor{darkblue}{rgb}{0.0, 0.25, 0.42}
\definecolor{blue}{rgb}{0.0,0.0,1.0}
\definecolor{green}{rgb}{0.0,1.0,0.0}

\begin{document}

\title{MOA-2022-BLG-249Lb: Nearby microlensing super-Earth planet detected from high-cadence surveys}
\titlerunning{MOA-2022-BLG-249Lb: Nearby microlensing super-Earth planet}

\author{
     Cheongho~Han\inst{01}
\and Andrew~Gould\inst{02,03}
\and Youn~Kil~Jung\inst{04,05}
\and Ian~A.~Bond\inst{06}
\and Weicheng~Zang\inst{07,08}
\\
(Leading authors)\\
     Sun-Ju~Chung\inst{04, 07}
\and Michael~D.~Albrow\inst{09}
\and Kyu-Ha~Hwang\inst{04}
\and Yoon-Hyun~Ryu\inst{04}
\and In-Gu~Shin\inst{07}
\and Yossi~Shvartzvald\inst{10}
\and Hongjing~Yang\inst{08}
\and Jennifer~C.~Yee\inst{07}
\and Sang-Mok~Cha\inst{04,11}
\and Doeon~Kim\inst{01}
\and Dong-Jin~Kim\inst{04}
\and Seung-Lee~Kim\inst{04}
\and Chung-Uk~Lee\inst{04}
\and Dong-Joo~Lee\inst{04}
\and Yongseok~Lee\inst{04,11}
\and Byeong-Gon~Park\inst{04}
\and Richard~W.~Pogge\inst{03}
\\
(The KMTNet collaboration)\\
     Shude~Mao \inst{08}
\and Wei~Zhu \inst{08}
\\
(Microlensing Astronomy Probe Collaboration)\\
     Fumio~Abe\inst{12}
\and Richard~Barry\inst{13}
\and David~P.~Bennett\inst{13,14}
\and Aparna~Bhattacharya\inst{13,14}
\and Hirosame~Fujii\inst{12}
\and Akihiko~Fukui\inst{15,16}
\and Ryusei~Hamada\inst{17}
\and Yuki~Hirao\inst{17}
\and Stela~Ishitani Silva\inst{14,18}
\and Yoshitaka~Itow\inst{12}
\and Rintaro~Kirikawa\inst{17}
\and Iona~Kondo\inst{17}
\and Naoki~Koshimoto\inst{19}
\and Yutaka~Matsubara\inst{12}
\and Sho~Matsumoto\inst{17}
\and Shota~Miyazaki\inst{17}
\and Yasushi~Muraki\inst{12}
\and Arisa~Okamura\inst{17}
\and Greg~Olmschenk\inst{13}
\and Cl{\'e}ment~Ranc\inst{20}
\and Nicholas~J.~Rattenbury\inst{21}
\and Yuki~Satoh\inst{17}
\and Takahiro~Sumi\inst{17}
\and Daisuke~Suzuki\inst{17}
\and Taiga~Toda\inst{17}
\and Mio~Tomoyoshi\inst{17}
\and Paul~J.~Tristram\inst{22}
\and Aikaterini~Vandorou\inst{13,14}
\and Hibiki~Yama\inst{17}
\and Kansuke~Yamashita\inst{17}
\\
(The MOA Collaboration)\\
}

\institute{
      Department of Physics, Chungbuk National University, Cheongju 28644, Republic of Korea,                                                            
\and  Max-Planck-Institute for Astronomy, K\"{o}nigstuhl 17, 69117 Heidelberg, Germany                                                                   
\and  Department of Astronomy, Ohio State University, 140 W. 18th Ave., Columbus, OH 43210, USA                                                          
\and  Korea Astronomy and Space Science Institute, Daejon 34055, Republic of Korea                                                                       
\and  Korea University of Science and Technology, Korea, (UST), 217 Gajeong-ro, Yuseong-gu, Daejeon, 34113, Republic of Korea                            
\and  Institute of Natural and Mathematical Science, Massey University, Auckland 0745, New Zealand                                                       
\and  Center for Astrophysics $|$ Harvard \& Smithsonian, 60 Garden St., Cambridge, MA 02138, USA                                                        
\and  Department of Astronomy, Tsinghua University, Beijing 100084, China                                                                                
\and  University of Canterbury, Department of Physics and Astronomy, Private Bag 4800, Christchurch 8020, New Zealand                                    
\and  Department of Particle Physics and Astrophysics, Weizmann Institute of Science, Rehovot 76100, Israel                                              
\and  School of Space Research, Kyung Hee University, Yongin, Kyeonggi 17104, Republic of Korea                                                          
\and  Institute for Space-Earth Environmental Research, Nagoya University, Nagoya 464-8601, Japan                                                        
\and  Code 667, NASA Goddard Space Flight Center, Greenbelt, MD 20771, USA                                                                               
\and  Department of Astronomy, University of Maryland, College Park, MD 20742, USA                                                                       
\and  Department of Earth and Planetary Science, Graduate School of Science, The University of Tokyo, 7-3-1 Hongo, Bunkyo-ku, Tokyo 113-0033, Japan      
\and  Instituto de Astrof{\'i}sica de Canarias, V{\'i}a L{\'a}ctea s/n, E-38205 La Laguna, Tenerife, Spain                                               
\and  Department of Earth and Space Science, Graduate School of Science, Osaka University, Toyonaka, Osaka 560-0043, Japan                               
\and  Department of Physics, The Catholic University of America, Washington, DC 20064, USA                                                               
\and  Department of Astronomy, Graduate School of Science, The University of Tokyo, 7-3-1 Hongo, Bunkyo-ku, Tokyo 113-0033, Japan                        
\and  Sorbonne Universit\'e, CNRS, UMR 7095, Institut d'Astrophysique de Paris, 98 bis bd Arago, 75014 Paris, France                                     
\and  Department of Physics, University of Auckland, Private Bag 92019, Auckland, New Zealand                                                            
\and  University of Canterbury Mt.~John Observatory, P.O. Box 56, Lake Tekapo 8770, New Zealand                                                          
}


\abstract
{}
{
We investigate the data collected by the high-cadence microlensing surveys during the 2022
season in search for planetary signals appearing in the light curves of microlensing events.
From this search, we find that the lensing event MOA-2022-BLG-249 exhibits a brief positive 
anomaly that lasted for about 1 day with a maximum deviation of $\sim 0.2$~mag from a 
single-source single-lens model.
}
{
We analyze the light curve under the two interpretations of the anomaly: one originated by a
low-mass companion to the lens (planetary model) and the other originated by a faint
companion to the source (binary-source model).
}
{
It is found that the anomaly is better explained by the planetary model than the binary-source
model.  We identify two solutions rooted in the inner--outer degeneracy, for both of which  
the estimated planet-to-host mass ratio, $q\sim 8\times 10^{-5}$, is very small.  With the 
constraints provided by the microlens parallax and the lower limit on the Einstein radius, as 
well as the blend-flux constraint, we find that the lens is a planetary system, in which a 
super-Earth planet, with a mass $(4.83\pm 1.44)~M_\oplus$, orbits a low-mass host star, with 
a mass $(0.18\pm 0.05)~M_\odot$, lying in the Galactic disk at a distance $(2.00\pm 0.42)$~kpc.  
The planet detection demonstrates the elevated microlensing sensitivity of the current 
high-cadence lensing surveys to low-mass planets.
}
{}


\maketitle

\section{Introduction}\label{sec:one}

The microlensing method of finding planets has various advantages that can complement other
planet detection methods.  Especially, it provides a unique tool to detect planets belonging 
to faint stars because the lensing characteristics do not depend on the light of a lensing object. 
The method is also useful in detecting outer planets because of the high microlensing sensitivity 
to planets lying at around the Einstein radius which approximately corresponds to the snow line 
of a planetary system. See the review paper of \citet{Gaudi2012} for the discussion of various
advantages of the microlensing method.

\begin{table*}[t]
\small
\caption{Low-mass microlensing planets\label{table:one}}
\begin{tabular}{lllll}
\hline\hline
\multicolumn{1}{c}{Planet}               &
\multicolumn{1}{c}{Type}                 &
\multicolumn{1}{c}{Reference}            \\
\hline
OGLE-2005-BLG-390Lb       &   super-earth                      &   \citet{Beaulieu2006}               \\
MOA-2007-BLG-192Lb        &   super-earth                      &   \citet{Bennett2008}                \\
MOA-2009-BLG-266Lb        &   super-earth                      &   \citet{Muraki2011}                 \\
MOA-2011-BLG-262Lb        &   super-earth                      &   \citet{Bennett2014}                \\
MOA-2013-BLG-605Lb        &   super-earth                      &   \citet{Sumi2016}                   \\
OGLE-2013-BLG-0341Lb      &   terrestrial planet               &   \citet{Gould2014}                  \\
OGLE-2016-BLG-1195Lb      &   Earth-mass planet                &   \citet{Shvartzvald2017}, \citet{Bond2017}, \citet{Vandorou2023}  \\
OGLE-2016-BLG-1928L       &   terrestrial-mass rogue planet    &   \citet{Mroz2020}                   \\
OGLE-2017-BLG-0482Lb      &   super-earth                      &   \citet{Han2018}                    \\
OGLE-2017-BLG-1806Lb      &   super-earth                      &   \citet{Zang2023}                   \\
KMT-2017-BLG-0428Lb       &   super-earth                      &   \citet{Zang2023}                   \\
KMT-2017-BLG-1003Lb       &   super-earth                      &   \citet{Zang2023}                   \\
KMT-2017-BLG-1194Lb       &   super-earth                      &   \citet{Zang2023}                   \\
OGLE-2018-BLG-0532Lb      &   super-earth                      &   \citet{Ryu2020}                    \\
OGLE-2018-BLG-0677Lb      &   super-earth                      &   \citet{Herrera2020}                \\
OGLE-2018-BLG-0977Lb      &   super-earth                      &   \citet{Hwang2022}                  \\
OGLE-2018-BLG-1185Lb      &   super-earth                      &   \citet{Kondo2021}                  \\
KMT-2018-BLG-1025Lb       &   super-earth                      &   \citet{Han2021}                    \\
KMT-2018-BLG-1988Lb       &   super-earth                      &   \citet{Han2022a}                   \\
KMT-2018-BLG-0029Lb       &   super-earth                      &   \citet{Gould2020}                  \\
OGLE-2019-BLG-0960Lb      &   super-earth                      &   \citet{Yee2021}                    \\
OGLE-2019-BLG-1053Lb      &   terrestrial planet               &   \citet{Zang2021b}                  \\
KMT-2019-BLG-0253Lb       &   super-earth                      &   \citet{Hwang2022}                  \\
KMT-2019-BLG-1367Lb       &   super-earth                      &   \citet{Zang2023}                   \\
KMT-2019-BLG-1806Lb       &   super-earth                      &   \citet{Zang2023}                   \\
KMT-2020-BLG-0414Lb       &   Earth-mass planet                &   \citet{Zang2021a}                  \\
KMT-2021-BLG-0912Lb       &   super-earth                      &   \citet{Han2022b}                   \\
KMT-2021-BLG-1391Lb       &   super-earth                      &   \citet{Ryu2022}                    \\
\hline                                                 
\end{tabular}
\end{table*}

Another important advantage of the microlensing method is its high sensitivity to low-mass
planets \citep{Bennett1996}.  In general, the microlensing signal of a planet appears as a 
discontinuous perturbation to the smooth and symmetric lensing light curve produced by the 
host of the planet \citep{Mao1991, Gould1992b}. The amplitude of the planetary microlensing 
signal weakly depends on the planet-to-host mass ratio $q$, although the duration of the 
signal becomes shorter  in proportion to $q^{-1/2}$ \citep{Han2006}.  This implies that the 
microlensing sensitivity can extend to lower-mass planets as the observational cadence 
becomes higher.

The observational cadence of microlensing surveys has greatly enhanced during the 2010s with 
the replacement of the cameras installed on the telescopes of the previously established 
surveys of the Microlensing Observations in Astrophysics survey \citep[MOA:][]{Bond2001} 
and the Optical Gravitational Lensing Experiment \citep[OGLE:][]{Udalski2015} with new 
cameras having very wide field of views (FOVs) together with the commencement of a new 
survey of the Korea Microlensing Telescope Network \citep[KMTNet:][]{Kim2016}.  With the 
launch of these high-cadence surveys, lensing events can be observed with a cadence down 
to 0.25~hr compared to a 1~day cadence of earlier surveys.

The detection rate of very low-mass planets has greatly increased with the enhanced 
sensitivity to very short anomalies in microlensing light curves from this higher 
observational cadence.  In Table~\ref{table:one}, we list the discovered microlensing 
planets with masses below that of a super-Earth planet together with brief comments of 
the planet types and related references. Among these 28 planets, 5 are terrestrial planets 
with masses similar to that of Earth, and the other 23 are super-Earth planets with masses 
higher than Earth's but substantially below those of ice giants in the Solar System, that 
is, Uranus and Neptune.  To be noted is that 22 planets (79\%) have been detected since 
the full operation of the current high-cadence lensing surveys.  Very low-mass planets 
detected before the era of high-cadence survey were found using a specially designed 
observational strategy, in which survey groups focused on detecting lensing events and 
followup groups densely observed the events found by the survey groups with the employment 
of multiple narrow-FOV telescopes.  However, the detection rate of low-mass planets based 
on this strategy was low because of the limited number of events that could be observed by 
followup groups. In contrast, high-cadence surveys can densely monitor all lensing events 
without the need of extra followup observations.

In this paper, we report the discovery of a super-Earth planet found from inspection of 
the 2022 season microlensing data collected by the KMTNet and MOA surveys. The planet was 
discovered by analyzing the light curve of the microlensing event MOA-2022-BLG-249, for which 
a very short-term anomaly was covered by the survey data despite of its weak deviation. We 
check various interpretations of the signal and confirm its planetary origin.

We present our analysis according to the following organization. In Sect.~\ref{sec:two}, we 
describe the observations of the planetary lensing event and the data obtained from these 
observations. In Sect.~\ref{sec:three}, we depict the characteristics of the event and the 
anomaly appearing in the lensing light curve. We present the analyses of the light curve 
conducted under various interpretations of the anomaly and investigate higher-order effects 
that affect the lensing-magnification pattern.  We identify the source star of the event and 
check the feasibility of measuring the angular Einstein radius in Sect.~\ref{sec:four}, and 
estimate the physical parameters in Sect.~\ref{sec:five}.  We summarize the results of the 
analysis and conclude in Sect.~\ref{sec:six}.

\section{Observations and data}\label{sec:two}

The microlensing event MOA-2022-BLG-249 occurred on a source lying toward the Galactic bulge 
field at $({\rm RA},{\rm Dec})_{{\rm J2000}} = $ (17:55:27.73 -28:18:21.82), $(l,b) = 
(+1^\circ\hskip-2pt.65,-1^\circ\hskip-2pt.53)$.  The magnification of the source flux induced 
by lensing was first found by the MOA group on 2022 May 22, which corresponds to the abridged 
heliocentric Julian date of ${\rm HJD}^\prime \equiv {\rm HJD}- 2450000 = 9721.48$.  The KMTNet 
group identified the event at ${\rm HJD}^\prime =9721.63$, that is, 4 hours after the MOA 
discovery and designated the event as KMT-2022-BLG-0874. Hereafter, we refer to the event as 
MOA-2022-BLG-249 in accordance with the convention of the microlensing community using the 
event ID reference of the first discovery group.  The event lies approximately $100^{\prime\prime}$ 
outside the footprint of the OGLE survey. In any case, there are no data from the survey because 
the OGLE telescope was shut down during most of the 2022 season due to the Covid-19 pandemic. The 
source location corresponds to a sub-prime field of the MOA survey, and thus the coverage of the 
event is relatively sparse. In contrast, the source was in the KMTNet prime fields of BLG02 and 
BLG42, toward which observations were conducted with a high combined cadence of 0.25~hr, and thus 
the light curve of the event was densely covered by the KMTNet data.  The event was additionally 
observed by a survey of the Microlensing Astronomy Probe (MAP) collaboration, with a cadence of 
1-2 points per night.  The source flux gradually increased until the lensing light curve reached 
its peak on 2022 May 27 (${\rm HJD}^\prime \sim 9727$), and then returned to the baseline.  The 
duration of the event is very long, and the lensing magnification lasted throughout the whole 
2022 bulge season.

The event was observed with the use of multiple telescopes operated by the individual survey 
and followup groups. The MOA group utilized the 1.8~m telescope of the Mt.~John Observatory in 
New Zealand, the KMTNet group made the use of the three identical 1.6~m telescopes lying at the 
Siding Spring Observatory in Australia (KMTA), the Cerro Tololo Interamerican Observatory in 
Chile (KMTC), and the South African Astronomical Observatory in South Africa (KMTS), and the 
MAP group used the 3.6~m Canada-France-Hawaii Telescope (CFHT) in Hawaii. Data reduction 
and photometry of the event were done using the photometry pipelines of the individual groups, 
and the error bars of the individual data sets were readjusted using the routine described in 
\citet{Yee2012}.

\begin{figure}[t]
\includegraphics[width=\columnwidth]{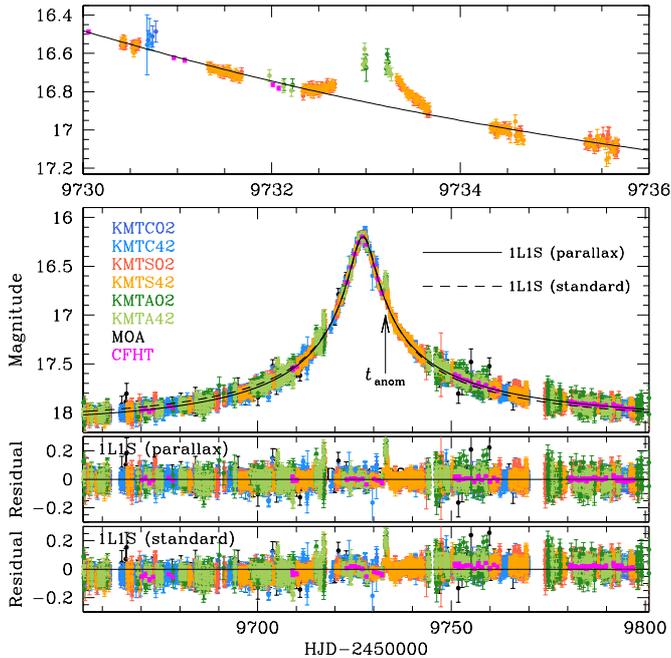}
\caption{
Light curve of the microlensing event MOA-2022-BLG-249.  The arrow marked by $t_{\rm anom}$ 
in the second panel indicates the location of the anomaly.  The top panel shows the enlarged 
view around the anomaly region.  The solid and dashed curves drawn over the data points are 
1L1S models obtained with (parallax model) and without (standard model) the consideration of 
microlens-parallax effects.  The two lower panels shows the residuals from the two models.
}
\label{fig:one}
\end{figure}

Figure~\ref{fig:one} shows the lensing light curve of MOA-2022-BLG-249.  The solid and dashed 
curves drawn over the data points are single-source single-lens (1L1S) models obtained from 
modeling with (parallax model) and without (standard model) the consideration of microlens-parallax 
effects \citep{Gould1992a}.  Detailed discussion on the parallax effects is presented in 
Sect.~\ref{sec:three}.  Although the observed light curve appears to be well described by the 
1L1S model, we find that there exists a brief anomaly appearing at $t_{\rm anom}\sim 9733$, 
which corresponds to about 6 days after the peak of the light curve. The upper panel of 
Figure~\ref{fig:one} shows the enlarged view of the region around the anomaly. The anomaly 
exhibited a positive deviation from the 1L1S model, and it lasted for about 1~day with a 
maximum deviation of $\Delta I \sim 0.2$~mag. The anomaly was mostly covered by the combination 
of the KMTS and KMTA data sets, and the region just before the major deviation was additionally 
covered by the two CFHT data points. The anomaly during the time gaps among the KMTS and KMTA 
coverage could have been covered by the KMTC data set, but the Chilean site was clouded out 
during the five day period around the time of the anomaly. Similarly, the MOA group did not 
observe this field during a 12-day interval that included the anomaly.

\begin{figure}[t]
\includegraphics[width=\columnwidth]{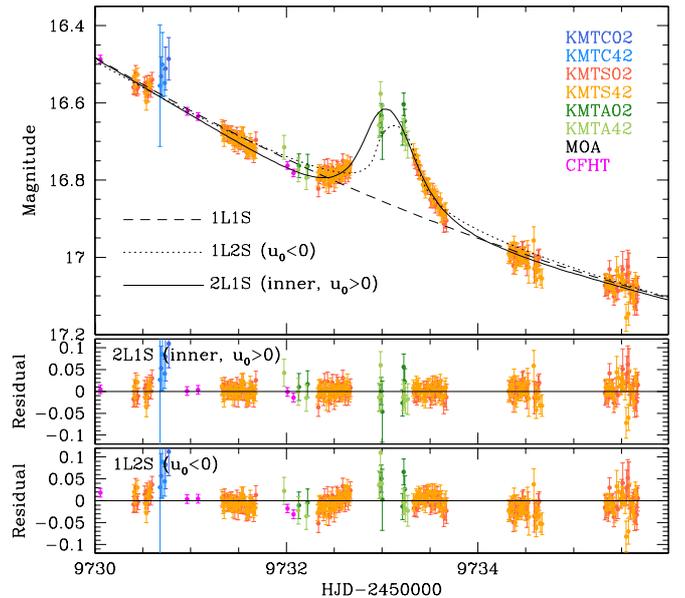}
\caption{
Comparison of the 2L1S and 1L2S models. The lower two panels show the residuals from the individual 
models.
}
\label{fig:two}
\end{figure}

\section{Light curve analysis}\label{sec:three}

It is known that a brief positive anomaly in a lensing light curve can arise via two channels: 
one by a low-mass companion to the lens \citep{Mao1991, Gould1992b} and the other by a faint 
companion to the source \citep{Gaudi1998}. In this section, we present the analysis of the lensing 
light curve conducted to reveal the nature of the anomaly. Details of the analysis based on the 
single-lens binary-source (1L2S) and the binary-lens and single-source (2L1S) interpretations are 
presented in the Sect.~\ref{sec:three-one} and ~\ref{sec:three-two}, respectively.

The analysis under each interpretation was carried out in search for a lensing solution, which 
represents a set of lensing parameters describing the observed lensing light curve. The common 
lensing parameters for both 2L1S and 1L2S models are $(t_0, u_0, \te, \rho)$, which represent 
the time of the closest lens-source approach, the projected lens-source separation scaled to 
the angular Einstein radius ($\thetae$) at $t_0$ (impact parameter), the event time scale, and
the source radius scaled to $\thetae$ (normalized source radius), respectively. Besides these 
basic parameters, a 2L1S modeling requires extra parameters of $(s, q, \alpha)$, where the first 
two parameters represent the projected separation (scaled to $\thetae$) and mass ratio between 
the lens components, respectively, and the last parameter denotes the angle of the source trajectory 
as measured from the binary-lens axis. A 1L2S model also requires additional parameters, including 
$(t_{0,2}, u_{0,2}, \rho_2, q_F)$, which represent the closest approach time, impact parameter and 
the normalized radius of the secondary source, and the flux ratio between the secondary and primary 
source stars, respectively.  In the 1L2S model, we designate the time of closest approach, the impact 
parameter and normalized radius of the primary source as $t_{0,1}$, $u_{0,1}$ and $\rho_1$, respectively, 
to distinguish them from those describing the secondary source.

In the modeling, we take the microlens-parallax effects into consideration because the event 
lasted for a significant fraction of a year. For a long time-scale event like MOA-2022-BLG-249, 
the deviation of the source motion from rectilinear caused by the orbital motion of Earth around 
the sun can be substantial \citep{Gould1992a}.  In order to consider these effects in the modeling, 
we add two extra lensing parameters $(\pien, \piee)$, which denote the north and east components of 
the microlens-lens parallax vector $\pivec_{\rm E} = (\pi_{\rm rel}/\thetae)(\muvec/\mu)$, respectively.  
Here $\muvec$ represents the vector of the relative lens-source proper motion and $\pi_{\rm rel}$ 
denotes the relative lens-source parallax, which is related to the distance to the lens, $\dl$, and 
source, $\ds$, by $\pi_{\rm rel} = {\rm AU}(1/\dl-1/\ds)$.  In each parallax modeling, we check a 
pair of solutions with $u_0 > 0$ and $u_0 < 0$.

\subsection{1L2S model}\label{sec:three-one}

The 1L2S modeling was carried via a downhill approach
using the Markov Chain Monte Carlo (MCMC) method because the lensing magnification smoothly 
changes with the variation of the 1L2S lensing parameters.  The initial parameters of $(t_0, 
u_0, \te)$ were given by adopting the values obtained from the 1L1S modeling, and those related 
to the source companion, that is, $(t_{0,2}, u_{0,2}, \rho_2, q_F)$, were given by considering 
the location and magnitude of the anomaly.  See \citet{Hwang2013} for details of the 1L2S 
modeling.  The lensing parameters of the $u_0>0$ and $u_0<0$ solutions and their $\chi^2$ 
values of the fit together with the degree of freedom (dof) are listed in Table~\ref{table:two}.  
It is found that the solution with $u_{0,1}<0$ results in a slightly better fit than the 
solution with $u_{0,1}>0$, by $\Delta\chi^2=1.6$.  The model curve of the $u_{0,1}<0$ solution 
and its residual in the region of the anomaly are shown in Figure~\ref{fig:two}.  It is found 
that the 1L2S models approximately delineate the observed anomaly, but they leave slight residual 
both in the rising and falling parts of the anomaly.  Especially, the negative residuals in the 
rising part of the anomaly appears both in the KMTS and CFHT data sets, suggesting that these 
residuals are likely to be real.

\begin{table}[t]
\small
\caption{1L2S model parameters\label{table:two}}
\begin{tabular*}{\columnwidth}{@{\extracolsep{\fill}}lcccc}
\hline\hline
\multicolumn{1}{c}{Parameter}      &
\multicolumn{1}{c}{$u_{0,1}>0$}    &
\multicolumn{1}{c}{$u_{0,1}<0$}    \\
\hline
$\chi^2/{\rm dof}$         &    9865.2/9800              &   9863.6/9800    \\
$t_{0,1}$ (HJD$^\prime$)   &    $9727.102 \pm 0.007 $    &   $ 9727.093 \pm 0.006$   \\
$t_{0,2}$ (HJD$^\prime$)   &    $9733.058 \pm 0.009 $    &   $ 9733.050 \pm 0.008$   \\
$u_{0,1}$ ($10^{-3}$)      &    $23.5 \pm 0.9       $    &   $ -22.1 \pm 0.8     $   \\
$u_{0,2}$ ($10^{-3}$)      &    $-0.7 \pm 0.2       $    &   $ -0.9 \pm 0.2      $   \\
$\te$ (days)               &    $143.36 \pm 4.11    $    &   $ 147.62 \pm 3.95   $   \\
$\rho_1$ ($10^{-3}$)       &    $20.35 \pm 1.75     $    &   $ 18.72 \pm 1.82    $   \\
$\rho_2$ ($10^{-3}$)       &    $1.54 \pm 0.09      $    &   $ 1.51 \pm 0.09     $   \\
$\pien$                    &    $-0.397 \pm 0.040   $    &   $ -0.488 \pm 0.044  $   \\
$\piee$                    &    $0.262 \pm 0.015    $    &   $ 0.251 \pm 0.015   $   \\
$q_F$                      &    $0.0048 \pm 0.0002  $    &   $ 0.0052 \pm 0.00021$   \\
\hline                                                      
\end{tabular*}
\tablefoot{ ${\rm HJD}^\prime = {\rm HJD}- 2450000$.  }
\end{table}

\begin{table*}[t]
\footnotesize
\caption{Parameters of 2L1S models (parallax only)\label{table:three}}
\begin{tabular}{l|ll|ll}
\hline\hline
\multicolumn{1}{c|}{Parameter}          &
\multicolumn{2}{c|}{Inner}              &
\multicolumn{2}{c}{Outer}               \\
\multicolumn{1}{c|}{ }                   &
\multicolumn{1}{c}{$u_0>0$}           &
\multicolumn{1}{c|}{$u_0<0$}          &
\multicolumn{1}{c}{$u_0>0$}           &
\multicolumn{1}{c}{$u_0<0$}           \\
\hline
$\chi^2/{\rm dof}$            &  9682.4/9799              &   9690.5/9799           &     9695.9/9799            &  9703.6/9799              \\
$t_0$ (HJD$^\prime$)          &  $9727.192 \pm 0.006$     &   $9727.196 \pm 0.006$  &     $9727.192 \pm 0.007$   &  $9727.195 \pm 0.006$     \\
$u_0$                         &  $0.023 \pm 0.001   $     &   $-0.023 \pm 0.001  $  &     $0.023 \pm 0.001   $   &  $-0.022 \pm 0.001  $     \\
$\te$ (days)                  &  $133.81 \pm 2.97   $     &   $131.90 \pm 2.95   $  &     $134.11 \pm 2.89   $   &  $133.82 \pm 2.75   $     \\
$s$                           &  $1.086 \pm 0.002   $     &   $1.091  \pm 0.002  $  &     $0.967 \pm 0.003   $   &  $0.961 \pm 0.002   $     \\
$q$ ($10^{-5}$)               &  $7.55 \pm 0.44     $     &   $8.94  \pm 0.52    $  &     $8.01 \pm 0.47     $   &  $9.0 \pm 0.50      $     \\
$\alpha$ (rad)                &  $3.638\pm 0.002    $     &   $-3.599  \pm 0.002 $  &     $3.638 \pm 0.002   $   &  $-3.599 \pm 0.002  $     \\
$\rho$ ($10^{-3}$)            &  $< 1.2             $     &   $< 1.2             $  &     $< 1.2             $   &  $< 1.2             $     \\
$\pien$                       &  $-0.491 \pm 0.038  $     &   $-0.561  \pm 0.043 $  &     $-0.465 \pm 0.039  $   &  $-0.548 \pm 0.0425 $     \\
$\piee$                       &  $0.260 \pm 0.016   $     &   $0.276  \pm 0.014  $  &     $0.267 \pm 0.014   $   &  $0.276 \pm 0.0146  $     \\
\hline                                                 
\end{tabular}
\end{table*}

\begin{figure}[t]
\includegraphics[width=\columnwidth]{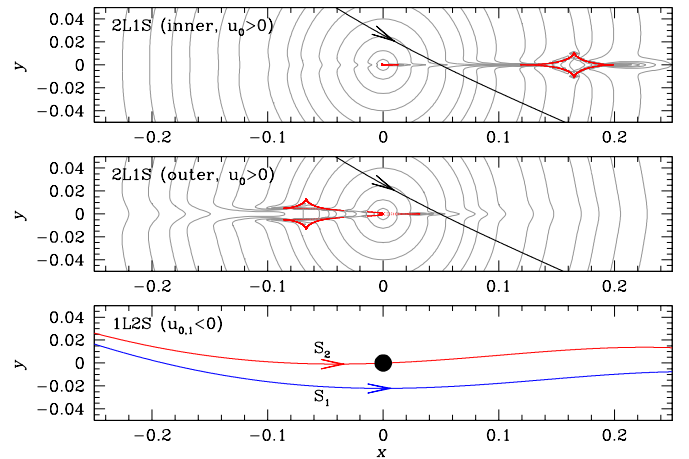}
\caption{
Lens system configurations. The two upper panels show the configurations of the inner and outer
2L1S solutions with $u_0 > 0$ and the bottom panel shows the configuration of the 1L2S solution 
with $u_{0,1}<0$. In each of the panels showing the 2L1S configurations, the red cuspy figures 
represent caustics, the line with an arrow is the source trajectory, and grey curves encompassing
 the caustic are equi-magnification contours. In the panel of the 1L2S solution, the black filled 
dot represent the lens, and the blue and red curves denote the trajectories of the primary (marked 
by $S_1$) and secondary (marked by $S_1$) source stars, respectively.
}
\label{fig:three}
\end{figure}

The lens-system configuration of the $u_{0,1}<0$ model is shown in the bottom panel of 
Figure~\ref{fig:three}, in which the arrowed curves marked in blue and red represent the 
trajectories of the primary (labeled as "$S_1$") and secondary source (labeled as "$S_2$") 
stars, respectively. According to the 1L2S interpretation, the anomaly was produced by the 
close approach of a secondary source to the lens. The secondary source is very faint, 
and its flux is $\sim 0.5\%$ of the flux from the primary source.

\subsection{2L1S model}\label{sec:three-two}

The 2L1S modeling was conducted in two steps. In the first step, we searched for the binary-lens
parameters $s$ and $q$ via a grid approach with multiple starting values of the source trajectory
angle $\alpha$, while we found the other lensing parameters via a downhill approach.
We then constructed a $\Delta\chi^2$ map on the $(s, q)$ parameter
plane and identified a pair of degenerate solutions resulting from the "inner-outer" degeneracy
\citep{Gaudi1997}. In the second step, we refined the lensing parameters of the individual local 
solutions by allowing all parameters to vary.

In Table~\ref{table:three}, we list the lensing parameters of the inner and outer 2L1S solutions, 
for each of which there are a pair of solutions with $u_0>0$ and $u_0<0$.  Among the solutions, it 
was found that the inner solution with $u_0>0$ yields the best fit to the data.  From the comparison 
of the 2L1S fit with that of 1L2S fit, it is found that the anomaly is better explained by the 2L1S 
interpretation than the 1L2S interpretation. In Figure~\ref{fig:two}, we draw the model curve of the 
inner 2L1S solution (with $u_0>0$) and its residual, showing that the residual of the 1L2S model 
around the anomaly does not appear in the residual of the 2L1S model. From the comparison of the 
fits, it is found that the 2L1S model provides a better fit to the data than the 1L2S model by 
$\Delta\chi^2=181.2$, indicating that the origin of the perturbation is a low-mass companion to 
the lens rather than a faint companion to the source.

\begin{figure}[t]
\includegraphics[width=\columnwidth]{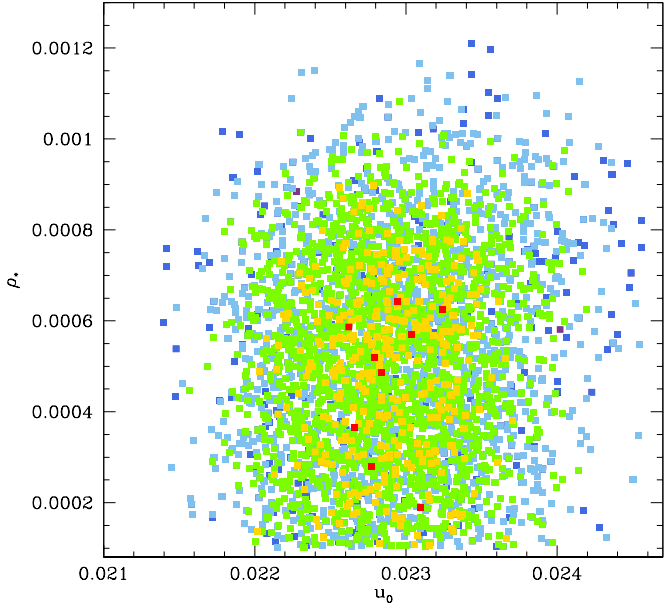}
\caption{
Scatter plot of points in the MCMC chain on the $(u_0, \rho)$ parameter plane obtained from 
the 2L1S modeling.  The color coding is set to designate points with $\leq 1\sigma$ (red), 
$\leq 2\sigma$ (yellow), $\leq 3\sigma$ (green), $\leq 4\sigma$ (cyan), and $\leq 5\sigma$ (blue).
}
\label{fig:four}
\end{figure}

The lens-system configurations of the inner and outer 2L1S solutions with $u_0>0$  values are 
shown in the two upper panels of Figure~\ref{fig:three}.  According to the inner and outer 
solutions, the anomaly was produced by the source passages through the regions lying on the 
side close to and farther from  the planetary caustic, respectively.  The inner and outer 
solutions can be viewed as "wide" and "close" solutions, respectively, arising due to the 
similarity between the centrals caustics induced by a wide planet and a close planet: "close-wide" 
degeneracy \citep{Griest1998}.  \citet{Yee2021} pointed out that the transition between the 
outer-inner and close-wide degeneracies is continuous, and \citet{Hwang2022} introduced an 
analytic expression for the relation between the binary separations of the inner ($s_{\rm in}$) 
and outer ($s_{\rm out}$) solutions:
\begin{equation}
s^\dagger = (s_{\rm in}\times s_{\rm out})^{1/2} =
{(u_{\rm anom}^2+4)^{1/2} + u_{\rm anom}  \over  2}.
\end{equation}
Here $u_{\rm anom}= (\tau_{\rm anom}^2+ u_0^2)^{1/2}$ represents the lens-source separation at the 
time of the anomaly $t_{\rm anom}$, and $\tau_{\rm anom}=(t_{\rm anom}-t_0)/\te$.  It is found that 
the value of $s^\dagger$ estimated from the planet separations $(s_{\rm in}, s_{\rm out})=(1.086, 
0.967)$, that is, $s^\dagger =(s_{\rm in} \times s_{\rm out})^{1/2}=1.024$, matches very well the 
value estimated from the lensing parameters $(t_0, u_0, \te, t_{\rm anom})$, that is, $s^\dagger = 
[(u_{\rm anom}^2 + 4)^{1/2} + u_{\rm anom}]/2 =1.024$. The estimated companion-to-primary mass ratio, 
$q\sim 8\times 10^{-5}$, is very low for both the inner and outer solutions, and the event time scale, 
$\te \sim 134$~days, is substantially longer than the several weeks of typical Galactic lensing events.  
The normalized source radius cannot be accurately measured because the source did not cross the caustic, 
and only the upper limit, $\rho_{\rm max} = 1.2\times 10^{-3}$, can be placed. See the scatter plot of 
the MCMC points on the $(u_0, \rho)$ parameter plane presented in Figure~\ref{fig:four}.

\begin{table*}[t]
\footnotesize
\caption{Parameters of 2L1S models (orbit+parallax)\label{table:four}}
\begin{tabular}{l|ll|ll}
\hline\hline
\multicolumn{1}{c|}{Parameter}          &
\multicolumn{2}{c|}{Inner}              &
\multicolumn{2}{c}{Outer}               \\
\multicolumn{1}{c|}{ }                   &
\multicolumn{1}{c}{$u_0>0$}           &
\multicolumn{1}{c|}{$u_0<0$}          &
\multicolumn{1}{c}{$u_0>0$}           &
\multicolumn{1}{c}{$u_0<0$}           \\
\hline
$\chi^2/{\rm dof}$            &    9682.1/9801            &   9690.0/9801             &      9682.3/9801              &   9686.1/9801               \\
$t_0$ (HJD$^\prime$)          &    $9727.191 \pm 0.006$   &   $9727.198 \pm 0.007$    &      $9727.188 \pm 0.006$     &   $9727.190 \pm 0.006$      \\
$u_0$                         &    $0.023 \pm 0.001   $   &   $-0.022 \pm 0.001  $    &      $0.023 \pm 0.001   $     &   $-0.022 \pm 0.001  $      \\
$\te$ (days)                  &    $134.50 \pm 2.78   $   &   $133.99 \pm 3.12   $    &      $135.56 \pm 3.14   $     &   $135.94 \pm 2.35   $      \\
$s$                           &    $1.073 \pm 0.009   $   &   $1.077 \pm 0.011   $    &      $0.942 \pm 0.007   $     &   $0.932 \pm 0.009   $      \\
$q$ ($10^{-5}$)               &    $7.22 \pm 0.72     $   &   $8.89 \pm 0.93     $    &      $7.17 \pm 0.69     $     &   $8.07 \pm 0.60     $      \\
$\alpha$ (rad)                &    $3.646 \pm 0.015   $   &   $-3.614 \pm 0.016  $    &      $3.642 \pm 0.015   $     &   $-3.615 \pm 0.010  $      \\
$\rho$ ($10^{-3}$)            &    $< 1.2             $   &   $< 1.2             $    &      $< 1.2             $     &   $< 1.2             $      \\
$\pien$                       &    $-0.464 \pm 0.037  $   &   $-0.557 \pm 0.045  $    &      $-0.474 \pm 0.038  $     &   $-0.529 \pm 0.040  $      \\
$\piee$                       &    $0.270 \pm 0.015   $   &   $0.276 \pm 0.015   $    &      $0.259 \pm 0.014   $     &   $0.276 \pm 0.014   $      \\
$ds/dt$ (yr$^{-1}$)           &    $0.705 \pm 0.54    $   &   $0.832 \pm 0.621   $    &      $1.70 \pm 0.42     $     &   $2.021 \pm 0.554   $      \\
$d\alpha/dt$ (yr$^{-1}$)      &    $-0.58 \pm 0.93    $   &   $0.964 \pm 1.009   $    &      $-0.29 \pm 0.94    $     &   $0.975 \pm 0.630   $      \\
\hline                                                 
\end{tabular}
\end{table*}

\begin{figure}[t]
\includegraphics[width=\columnwidth]{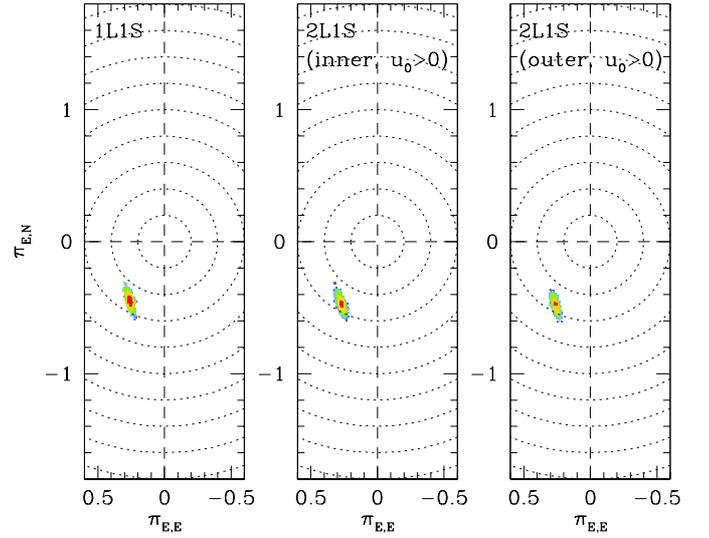}
\caption{
Scatter plots of the MCMC points in the chains of the 1L1S, and the inner and outer 2L1S 
solutions on the $(\piee, \pien)$ parameter plane.  In all cases, we present plots of the 
solutions with $u_0>0$, while the solutions with $u_0<0$ result in similar plots.  The dotted 
circles are drawn at every $0.2\pie$ interval. The color coding is same as that used in 
Fig.~\ref{fig:four}.
}
\label{fig:five}
\end{figure}

We note that the degeneracy between the 2L1S and 1L2S models was able to be securely resolved 
thanks to the nature of the event with a high peak magnification and an acute source trajectory 
angle.  In this case, the duration of the anomaly increases by a factor $|1/\sin \alpha|$ 
\citep{Yee2021}, which corresponds to a factor 2.1 in the case of MOA-2022-BLG-249.  That is, if 
this anomaly occurred with a right angle, that is, $\alpha \sim 90^\circ$ or $270^\circ$, then 
the anomaly would have been half as short, and the data might not have been good enough to 
distinguish the 2L1S model from the 1L2S model.  For events with acute trajectory angles, the 
magnification is lower than the peak magnification by a factor $|\sin \alpha|$.  
This factor is 0.47 (0.8 magnitudes) in the case of MOA-2022-BLG-249.

\subsection{Microlens-parallax effects}\label{sec:three-three}

It is found that considering parallax effects is important for the precise description of the observed 
light curve.  This is somewhat expected from the long time scale of the event. The improvement of fit 
with the parallax effect is huge, by $\Delta\chi^2=5300$ with respect to the model obtained under the 
assumption that the relative lens-source motion is rectilinear. The inner and outer 2L1S solutions 
result in similar values of the parallax parameters of $(\pien, \piee)\sim (-0.48, 0.27)$.

We checked the solidness of the parallax measurement by inspecting the consistency of the parallax 
parameters measured from the 1L1S and 2L1S modeling. Figure~\ref{fig:five} shows the scatter plots 
of MCMC points of the 1L1S solution and the inner and outer 2L1S solutions with positive $u_0$ 
values.  The parallax modeling was conducted by excluding the data around the perturbation 
$(9730<{\rm HJD}^\prime <9736)$ because the parallactic Earth motion has a long-term effect on 
the lensing light curve. From this check, it was found that all the tested models result in 
consistent parallax parameters, and this indicates that the parallax parameters are securely 
measured.

\begin{figure}[t]
\includegraphics[width=\columnwidth]{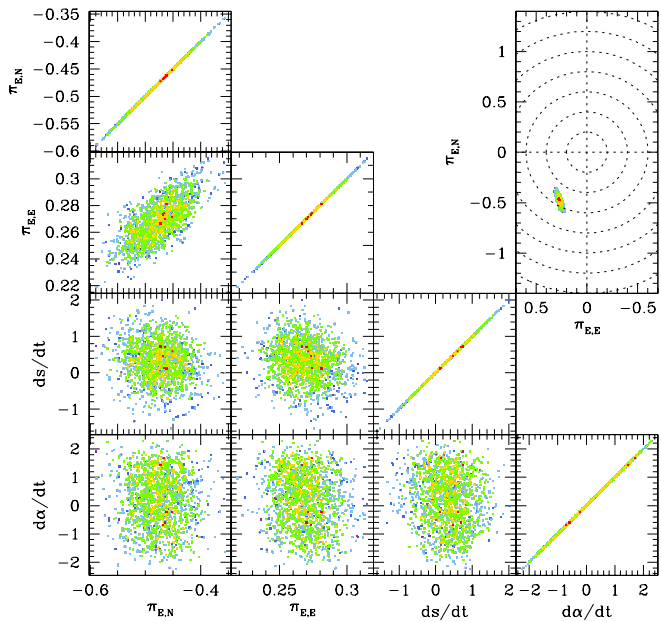}
\caption{
Scatter plot of the MCMC chain on the parameter planes of higher-order parameters of $(\pien, \piee, 
ds/dt, d\alpha/dt)$ for the inner 2L1S solution with $u_0>0$ obtained considering both microlens-parallax 
and lens-orbital effects.  The plot on the $(\piee, \pien)$ plane in the upper right inset is presented 
for the direct comparison with the plots presented in Fig.~\ref{fig:five}.
}
\label{fig:six}
\end{figure}

We also checked the effect of the planetary orbital motion on the $\pie$ measurement because the
planet might have moved during the 6~day period between the peak and the planetary perturbation 
and this could affect the lens system configuration. For this check, we tested additional models 
considering the planetary motion by including two orbital parameters of $(ds/dt, d\alpha/dt)$, 
which represent the annual change rates of the planetary separation and source trajectory angle, 
respectively.  The lensing parameters of the solutions considering the lens-orbital motion are 
listed in Table~\ref{table:four}. It is found that the lens-orbital motion does not have a 
significant effect on the microlens-parallax parameters.  This can be seen in Figure~\ref{fig:six}, 
in which we present the scatter plots of the MCMC points on the $(\pien, \piee, ds/dt, d\alpha/dt)$ 
parameter planes for the inner $u_0>0$ solution considering both the microlens-parallax and  
lens-orbital effects.  The plots show that the uncertainties of the orbital parameters, that is, 
$ds/dt$ and $d\alpha/dt$, are very large.  Although there are some variations of the plots in the 
orbital-parameter space for the other solutions, that is, the inner solution with $u_0<0$ and outer 
solutions with $u_0>0$ and $u_0<0$, the variation of the parallax parameters is minor.  Furthermore, 
the parallax parameters are similar to the values determined without considering the lens-orbital 
effect.  These results indicate that the effect of the lens-orbital motion on the light curve is minor.

We additionally checked the possibility that the parallax effect is imitated by the orbital effect 
induced by a source companion for which its luminosity contribution to the lensing light curve is 
negligible: xallarap effects \citep{Griest1992, Han1997, Smith2002}.  For this check, we conducted 
an additional modeling with the consideration of xallarap effects. Following the parameterization of 
\citep{Dong2009}, the xallarap modeling was done by including 5 extra parameters of $(\xi_{{\rm E},N},
\xi_{{\rm E},E}, P, \psi, i)$. Here the first two parameters $(\xi_{{\rm E},N},\xi_{{\rm E},E})$ denote 
the north and east components of the xallarap vector $\xivec_{\rm E}$, respectively, and the other 
parameters represent the period, phase angle, and inclination of the binary-source orbit, respectively. 
The magnitude of the xallarap vector, $\xi_{\rm E} = (\xi^2_{{\rm E},N}+\xi^2_{{\rm E},E})^{1/2}$, is 
related to the semi-major axis, $a$, of the source orbit by $\xi_{\rm E} =a_{\rm S}/\hat{r}_{\rm E}$, 
where $\hat{r}_{\rm E}=\ds\thetae$ denotes the physical Einstein radius projected onto the source 
plane, $a_{\rm S}=aM_{{\rm S},2}/(M_{{\rm S},1}+M_{{\rm S},2})$, and $(M_{{\rm S},1}, M_{{\rm S},2})$ 
are the masses of source components. Combined with the Kepler's law, the mass ratio between the source 
stars, $Q=M_{{\rm S},2}/M_{{\rm S},1}$, follows the relation \citep{Dong2009}
\begin{equation}
R={Q^3 \over (1+Q)^2} =
{(a_{\rm S}/{\rm AU})^3 \over (P/{\rm yr})^2 (M_{{\rm S},1}/M_\odot)}.
\label{eq1}
\end{equation}

\begin{figure}[t]
\includegraphics[width=\columnwidth]{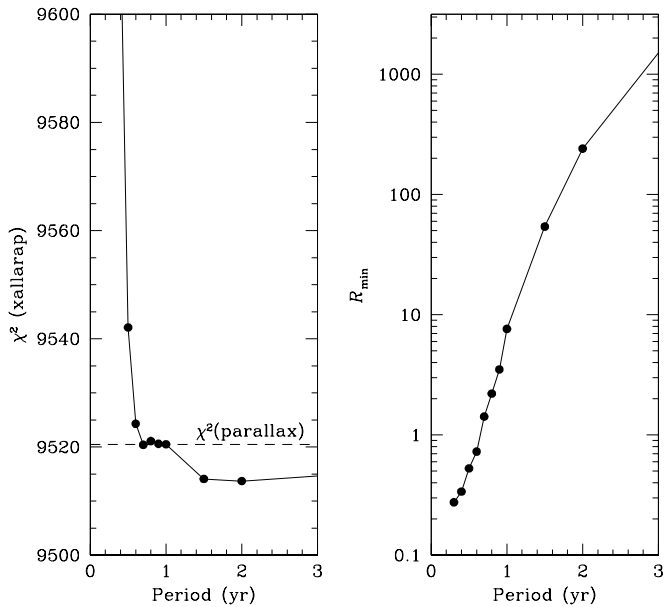}
\caption{
Results of xallarap modeling. The left panel shows the $\chi^2$ values of the xallarap fits as a 
function of the source orbital period, and the right panel shows the lower limit of $R=Q^3/(1+Q)^2$ 
as a function of the period. The dashed horizontal line in the left panel indicates the $\chi^2$ 
value of the parallax fit.
}
\label{fig:seven}
\end{figure}

The result of the xallarap modeling is presented in Figure~\ref{fig:seven}, in which the left and 
right panels show $\chi^2$ value of the xallarap fit and the lower limit of $R = Q^3/(1+Q)^2$ 
value with respect to the orbital period of the source, respectively.  For the computation of $R$, 
we adopted the mass of the primary source of $M_{S,1}\sim 1~M_\odot$ and distance to the source 
of $\ds=8$~kpc.  For the angular Einstein radius, we adopted the lower limit of $\theta_{\rm E,min}
\sim 0.46$~mas because $R \propto a_{\rm S} \propto \thetae$, and thus the lower limit of the $R$ 
value results from the lower limit of $\thetae$. The procedure of $\theta_{\rm E,min}$ determination 
is discussed in Sect.~\ref{sec:four}. From the comparison of the $\chi^2$ values between the xallarap, 
$\chi^2_{\rm xallarap}$, and parallax, $\chi^2_{\rm parallax}$, solutions, it is found that 
$\chi^2_{\rm xallarap}$ is higher than $\chi^2_{\rm parallax}$ for solutions with $P < 1$~yr, 
almost same as $\chi^2_{\rm parallax}$ for solutions with $P \sim 1$~yr, and slightly lower 
than $\chi^2_{\rm parallax}$ for solutions with $P > 1$~yr. For the solutions with $P > 1$~yr, 
the $\chi^2$ difference $\Delta\chi^2 = \chi^2_{\rm parallax} - \chi^2_{\rm xallarap} \lesssim 
6.8$ is very minor with 3 additional dof, and this corresponds to about 11\% probability even 
assuming Gaussian statistics. Furthermore, the ratio $R \gtrsim 40$ for these solutions, and 
thus the mass ratio ratio is $Q \gtrsim 40$, implying that mass of the source companion is 
$M_{{\rm S},2}\sim 40~M_\odot$, which corresponds to that of a black hole and thus unphysical. 
The results of the xallarap models indicate that there is no evidence that the light curve is 
affected by xallarap effects, and, more importantly, no evidence that the parallax signal is 
actually due to systematics.  Therefore, we conclude that the parallax signal is real.  As 
discussed in Sect.~\ref{sec:four}, the source may be a disk star lying in front of the bulge, 
and the source distance may be smaller than the adopted value of 8~kpc, but this has little 
impact on your conclusions from the xallarap modeling.

\section{Source star and Einstein radius}\label{sec:four}

In this section, we define the source star of the lensing event not only for fully characterizing 
the event but also for constraining the lensing observable of the angular Einstein radius. The 
value of $\thetae$ is estimated from the normalized source radius and angular source radius 
$\theta_*$ by $\thetae=\theta_*/\rho$.  Although the $\rho$ value cannot be measured for 
MOA-2022-BLG-249 due to the non-caustic-crossing nature of the anomaly, it is possible to constrain 
its upper limit, which yields the lower limit of the Einstein radius, that is, $\theta_{\rm E,min}
=\theta_*/\rho_{\rm max}$.

We specified the type of the source star by measuring its color and magnitude. For this
specification, we first placed the source in the instrumental color-magnitude diagram
(CMD) of stars around the source by measuring the $V$- and $I$-band magnitudes of the source
by regressing the light curve data measured in the individual passbands with respect to the 
lensing magnification estimated by the model. We then calibrated the source color and magnitude 
using the centroid of the red giant clump (RGC), for which its extinction-corrected (de-reddened) 
color and magnitude are known, as a reference \citep{Yoo2004}, that is,
\begin{equation}
(V-I, I)_{0,{\rm S}} = (V-I, I)_{0,{\rm RGC}} + [(V-I, I)_{\rm S} - (V-I, I)_{\rm RGC}].
\label{eq3}
\end{equation}
Here $(V-I, I)_{\rm S}$ and $(V-I, I)_{\rm RGC}$ denote the instrumental colors and magnitudes of 
the source and RGC centroid, respectively, and $(V-I, I)_{0,{\rm S}}$ and $(V-I, I)_{0,{\rm RGC}}$ 
indicate their corresponding de-reddened values.

\begin{figure}[t]
\includegraphics[width=\columnwidth]{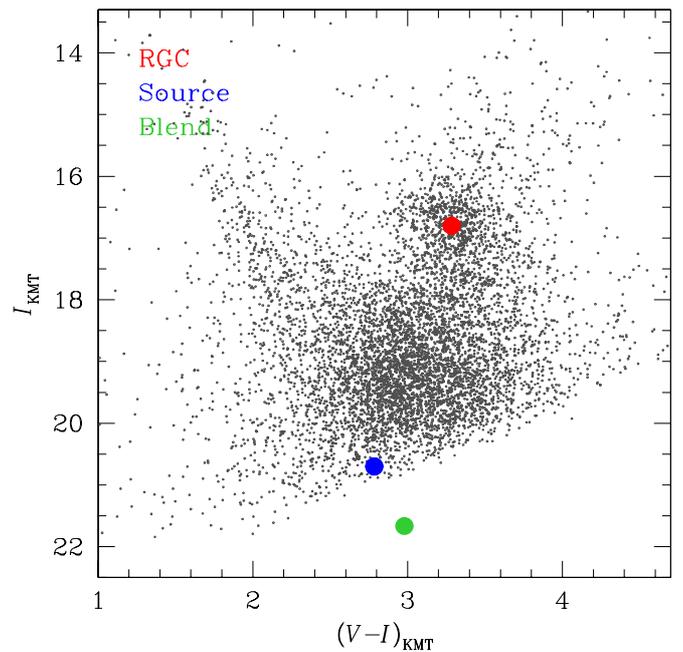}
\caption{
Source location with respect to the red giant clump (RGC) in the
instrumental CMD. Also marked is the location of the blend.
}
\label{fig:eight}
\end{figure}

Figure~\ref{fig:eight} shows the locations of the source and RGC centroid in the instrumental 
CMD. Also marked is the location of the blend. The instrumental color and magnitude are 
$(V-I, I)_{\rm S} = (2.785 \pm 0.017, 20.700 \pm 0.001)$ for the source and 
$(V-I, I)_{\rm RGC} = (3.285, 16.800)$ for the RGC centroid. With the known de-reddened color 
and magnitude of the RGC centroid, $(V-I, I)_{0,{\rm RGC}} = (1.060, 14.530)$ \citep{Bensby2013, 
Nataf2013}, we estimated the de-reddened color and magnitude of the source of
\begin{equation}
(V-I, I)_{0,{\rm S}} = (0.560 \pm 0.017, 18.430 \pm 0.001). 
\label{eq4}
\end{equation}
According to the estimated color and magnitude, the source is a mid to late F-type 
main-sequence star, and it probably lies in the disk in front of the bulge, although it could 
be a rare star in the bulge.

We also estimated the de-reddened color and brightness of the blend as $(V-I, I)_{0,b} = (0.76, 
19.31)$ assuming that the blend lies behind most of the dust, that is, in or near the bulge.  
We checked the possibility that the lens is the major source of the blended flux.  For this 
check, we measured the astrometric offset between the centroids of the source measured before 
and at the time of the lensing magnification. Considering that this offset is measured in the 
same season, it is expected that the offset would be smaller than the measurement uncertainty 
if the lens is the blend.  The measured offset in the KMTC image is $\Delta\theta = (169.2 \pm 
44.7)$~mas. This 3.8$\sigma$ offset is confirmed by the offset $\Delta\theta =(80\pm 10)$~mas 
measured in the CFHT data taken with seeing of 0.45$^{\prime\prime}$--0.55$^{\prime\prime}$.  
This indicates that the blend is caused by a nearby star lying close to the source rather than 
the lens.

With the specification of the source, we then estimate the angular radius of the source. 
For this, we first converted the measured $V-I$ color into $V-K$ color with the use of the 
\citet{Bessell1988} relation, and then estimated the angular radius of the source using the 
$(V-K, V)$--$\theta_*$ relation of \citet{Kervella2004}. We estimated that the source has 
an angular radius of
\begin{equation}
\theta_* = 0.551 \pm 0.040~\mu{\rm as},
\label{eq5}
\end{equation}
and this yields the minimum values of the angular Einstein radius 
\begin{equation}
\theta_{\rm E,min} = {\theta_* \over \rho_{\rm max}} = 0.46~{\rm mas},
\label{eq6}
\end{equation}
and the relative lens-source proper motion 
\begin{equation}
\mu_{\rm min} = { \theta_{\rm E,min} \over \te} = 1.25~{\rm mas/yr}.
\label{eq7}
\end{equation}

\section{Physical lens parameters}\label{sec:five}
   
The physical parameters of a lens are constrained by measuring the lensing observables of an
event. These observables include the event time scale $\te$, Einstein radius $\thetae$, and 
microlens parallax vector $\pivec_{\rm E}=(\pien,\piee)$, and the mass and distance to the 
lens are determined from the combination of these observables as
\begin{equation}
M={\thetae \over \kappa \pie};\qquad
\dl = {{\rm AU} \over \pie\thetae + \pi_{\rm S}},
\label{eq8}
\end{equation}
respectively \citep{Gould2000}. Here $\kappa=4G/(c^2{\rm AU})$ and $\pi_{\rm S}={\rm AU}/\ds$ 
is the parallax of the source.  For MOA-2022-BLG-249, the values of $\te$ and $\pie$ are 
securely measured, but the value of $\thetae$ cannot be measured and only its lower limit is 
constrained, making it difficult to analytically estimate $M$ and $\dl$ using the relations 
in Equation~(\ref{eq8}). We, therefore, estimate the physical lens parameters by conducting 
a Bayesian analysis based on the measured lensing observables and other available constraints.

In the first step of the Bayesian analysis, we generated a large number ($2 \times 10^8$) of 
artificial lensing events, for which the locations the lens and source and their relative 
proper motion were assigned on the basis of a Galactic model and the lens masses were allocated 
on the basis of a mass-function model by conducting a Monte Carlo simulation. In the simulation, 
we adopted the models of the Galaxy and lens mass function described in \citet{Jung2021} and 
\citet{Jung2018}, respectively. For each simulated event, we computed the lensing observables 
corresponding to the values of $M$, $\dl$, $\ds$, and $\mu$ by
\begin{equation}
\te={\thetae \over \mu};\qquad
\thetae=(\kappa M \pi_{\rm rel})^{1/2};\qquad
\pie={\pi_{\rm rel} \over \thetae}.
\label{eq9}
\end{equation}
In the second step, we imposed a weight $w_i = \exp(-\chi^2/2)$ to each artificial event and 
constructed posteriors of $M$ and $\dl$. 
In this procedure, the $\chi^2$ value is calculated as
\begin{equation}
\chi^2 =
\biggl({t_{\e,i} - t_\e\over \sigma_{t_\e}}\biggr)^2 +
\sum_{j,k=1}^2 b_{jk}(\pi_{\e,j,i}-\pi_{\e,i})(\pi_{\e,k,i}-\pi_{\e,i}),
\label{eq10}
\end{equation}
where $(\pi_{E,1},\pi_{E,2})_i=(\pien,\piee)_i$ is expressed in two component form,
$(t_{{\rm E},i},\pivec_{{\rm E},i})$ are the observables of each simulated event,
$(t_{\rm E},\pivec_{\rm E})$ represent the measured observables, $\sigma_{\te,}$
is the uncertainty in the $t_\e$ measurement, and $b_{jk}$ is the inverse covariance matrix 
of $\pivec_{\rm E}$. See, Equations (10) and (11) of \citet{Gould2022}. Finally, we 
imposed the constraint of the Einstein radius by setting $w_i = 0$ for events with $\thetae 
\leq \theta_{{\rm E},\rm min}$.

\begin{figure}[t]
\includegraphics[width=\columnwidth]{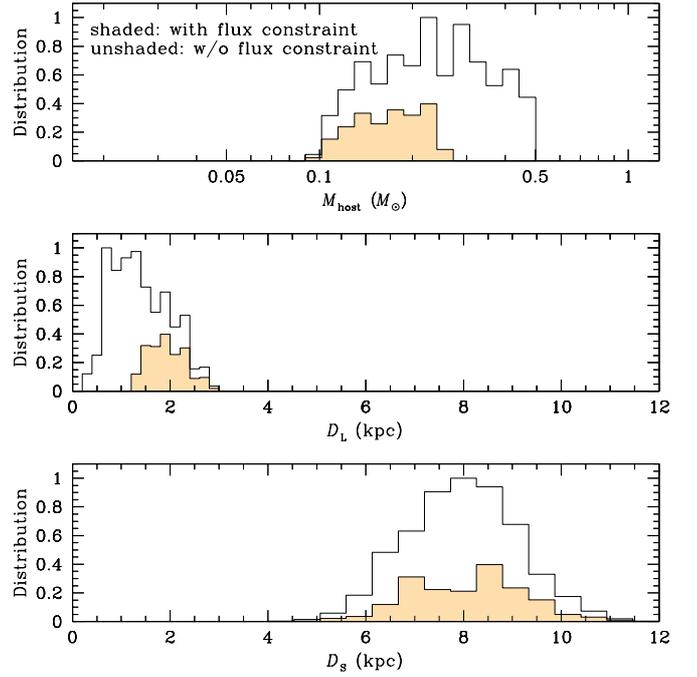}
\caption{
Bayesian posteriors of the lens mass, distance to the lens and source. In each panel, the 
shaded and and unshaded distributions are obtained with (shaded) and without (unshaded) 
imposing the blend-flux constraint, respectively.
}
\label{fig:nine}
\end{figure}

We imposed an additional constraint provided by the fact that the flux from the lens cannot 
be greater than the blend flux. Imposing this blend-flux constraint may be important because 
the distance to the lens expected from the large value of the measured microlens parallax, 
$\pie = (\pi^2_{{\rm E},N}+\pi^2_{{\rm E},E})^{1/2}\sim 0.55$, is small. In order to impose 
this constraint, we computed the lens magnitude as
\begin{equation}
I_{\rm L} = M_{I,{\rm L}} + 5 \log\left({\dl \over {\rm pc}}\right) - 5 + A_{I,{\rm L}}, 
\label{eq11}
\end{equation}
where $M_{I,{\rm L}}$ is the absolute $I$-band magnitude of a star corresponding to the 
lens mass, and $A_{I,{\rm L}}$ represents the $I$-band extinction to the lens. For the 
computation of $A_{I,{\rm L}}$, we modeled the extinction as
\begin{equation}
A_{I,{\rm L}} = A_{I,{\rm tot}}  \left[1 - \exp\left(-{|z| \over h_{z,{\rm dust}}}\right) 
\right],
\label{eq12}
\end{equation}
where $A_{I,{\rm tot}} = 2.49$ is the total $I$-band extinction toward the field, 
$h_{z,{\rm dust}} = 100$~pc is the vertical scale height of dust, $z = \dl \sin b  + z_0$, $b$ 
is the Galactic latitude, and $z_0=15$~pc is vertical position of the Sun above the Galactic 
plane.

Figure~\ref{fig:nine} shows the Bayesian posteriors of the host mass (top panel), $M_{\rm host}$, 
distances to the planetary system (middle panel) and source (bottom panel). We present two sets 
of posterior: one with (shaded distribution) and the other without (unshaded distribution) the 
blend-flux constraint. The posterior distributions show that the physical parameters are tightly 
defined despite the limited information on the angular Einstein radius. In Table~\ref{table:five}, 
we summarize the estimated physical parameters, in which the median values are presented as 
representative values and the uncertainties are estimated as 16\% and 84\% of the Bayesian posterior 
distributions. Here the planet mass is estimated as $M_{\rm planet}=qM_{\rm host}$, and the projected 
planet-host separation is computed by $a_\perp=s\thetae\dl$.  From the fact that the lower mass 
limit and the upper distance limit estimated using the analytic relations in Equation~(\ref{eq8}) 
based on the lower limit of the Einstein radius, that is, $M_{\rm min}=\theta_{\rm E,min}/
\kappa\pie\sim 0.12~M_\odot$ and $D_{\rm L,max}={\rm AU}/(\pie\theta_{\rm E,min}+\pi_{\rm S})\sim 
2.4$~kpc, match well the corresponding limits of the Bayesian posteriors indicates that these limits 
of the physical parameters are set by the combination of the constraints provided by 
$\theta_{\rm E,min}$ and $\pie$.  On the other hand, the upper limit of the mass and lower limit 
of the distance are set by the blend-flux constraint. This can be seen from the comparison of the 
Bayesian posteriors obtained with and without imposing the lens flux constraint.

\begin{table}[t]
\small
\caption{Physical lens parameters\label{table:five}}
\begin{tabular*}{\columnwidth}{@{\extracolsep{\fill}}lcccc}
\hline\hline
\multicolumn{1}{c}{Parameter}    &
\multicolumn{1}{c}{Inner}        &
\multicolumn{1}{c}{Outer}        \\
\hline
$M_{\rm host}$ ($M_\odot$)        &   $0.18 \pm 0.05$     & $\leftarrow$    \\
$M_{\rm planet}$ ($M_{\oplus}$)   &   $4.83 \pm 1.44 $    & $\leftarrow$    \\
$\dl$ (kpc)                       &   $2.00 \pm 0.42 $    & $\leftarrow$    \\
$a_\perp$ (AU)                    &   $1.63 \pm 0.35 $    & $1.45 \pm 0.31$ \\
\hline
\end{tabular*}
\end{table}

It turns out that the lens is a planetary system, in which a low-mass planet orbits a low-mass
host star lying in the Galactic disk. The estimated mass of the planet, $M_{\rm planet}\sim 
4.8~M_{\oplus}$, indicates that the planet is a super-Earth, and the detection of the system 
demonstrates the elevated microlensing sensitivity to low-mass planets with the increase of 
the observational cadence.  The estimated mass of the host, $M_{\rm host}\sim 0.18~M_\odot$, 
and distance, $\dl\sim 2.0$~kpc, indicate that the host of the planet is a very low-mass M dwarf 
lying in the Galactic disk.  Finding planets belonging to such low-mass stars using other methods 
is difficult because of the faintness of host stars, and thus the discovered planetary system well 
demonstrates the usefulness of the microlensing method in finding planets with low-mass host stars. 
The planetary system lies at a substantially closer distance than those of typical microlensing 
planets, which usually lie either in the bulge or in the portion of the disk that lies closer to 
the bulge than to the Sun.  We checked the hypothesis that the source is in the disk by additionally 
conducting a Bayesian analysis, in which we assumed that $\ds=7$~kpc and the dispersion of the 
source motion is negligible like that of disk stars.  We found that this analysis results in similar 
posteriors as those presented in Figure~\ref{fig:nine}, indicating that the uncertain source location 
has little effect on the result.

\section{Summary and conclusion}\label{sec:six}

We analyzed the microlensing event MOA-2022-BLG-249, for which the light curve exhibited 
a brief positive anomaly with a duration of  $\sim 1$~day and a maximum deviation of 
$\sim 0.2$~mag from a single-source single-lens model.  We tested both the planetary and 
binary-source origins, which are the two channels of producing a short-term positive anomaly 
in a lensing light curve.

We found that the anomaly was produced by a planetary companion to the lens rather than a 
binary companion to the source.  We identified two solutions rooted in the inner--outer 
degeneracy, for both of which  the estimated planet-to-host mass ratio, $q\sim 8\times 10^{-5}$, 
is very small.  With the constraints provided by the microlens parallax, the lower limit of 
the Einstein radius together with the blend-flux constraint, it was found that the lens is 
a planetary system, in which a super-Earth planet, with a mass $(4.83\pm 1.44)~M_\oplus$, 
orbits a low-mass host star, with a mass $(0.18\pm 0.05)~M_\odot$, lying in the Galactic 
disk at a distance $(2.00\pm 0.42)$~kpc.  The planet detection demonstrates the elevated 
microlensing sensitivity of the current high-cadence lensing surveys to low-mass planets.

\begin{acknowledgements}
Work by C.H. was supported by the grants of National Research Foundation of Korea 
(2020R1A4A2002885 and 2019R1A2C2085965).
This research has made use of the KMTNet system operated by the Korea Astronomy and Space 
Science Institute (KASI) and the data were obtained at three host sites of CTIO in Chile, 
SAAO in South Africa, and SSO in Australia.
This research was supported by the Korea Astronomy and Space Science Institute under the R\&D
program (Project No. 2023-1-832-03) supervised by the Ministry of Science and ICT.
The MOA project is supported by JSPS KAKENHI
Grant Number JSPS24253004, JSPS26247023, JSPS23340064, JSPS15H00781,
JP16H06287, and JP17H02871.
J.C.Y., I.G.S., and S.J.C. acknowledge support from NSF Grant No. AST-2108414. 
Y.S.  acknowledges support from NSF Grant No. 2020740.
This research uses data obtained through the Telescope Access Program (TAP), which has been
funded by the TAP member institutes. W.Zang, H.Y., S.M., and W.Zhu acknowledge support by the
National Science Foundation of China (Grant No. 12133005). W.Zang acknowledges the support
from the Harvard-Smithsonian Center for Astrophysics through the CfA Fellowship. W.Zhu
acknowledges the science research grants from the China Manned Space Project with No.\
CMS-CSST-2021-A11.
C.R. was supported by the Research fellowship of the Alexander von Humboldt Foundation.
\end{acknowledgements}

\end{document}